\documentclass[aps,prb,groupedaddress,showpacs,floatfix,twocolumn]{revtex4}%
\usepackage{amsmath,amssymb}
\usepackage{graphicx}
\usepackage{dcolumn}
\usepackage{enumerate}
\usepackage{comment}
\usepackage{color}
\newcommand{\rmd}{\mathrm{d}}
\newcommand{\dMdB}{$\rmd M/\rmd B$}
\newcommand{\dmdb}{$\rmd m/\rmd b_1$}
\newcommand{\rJJ}{$J^{(2)}{/}J^{(1)}$}
\newcommand{\rJ}[1]{$J_{#1}{/}J_1$}
\newcommand{\nmax}{$n_{\mathrm{max}}$}
\newcommand{\muB}{\mu_\mathrm{B}}
\newcommand{\kB}{k_\mathrm{B}}
\newcommand{\J}[1]{$J_{#1}$}
\newcommand{\JJ}[1]{$J^{({#1})}$}

\newcommand{\Fig}[1]{Fig.~\ref{Fig#1}}
\newcommand{\Figure}[1]{Figure~\ref{Fig#1}}
\newcommand{\Figs}[2]{Figs.~\ref{Fig#1} and \ref{Fig#2}}
\newcommand{\Figures}[2]{Figures~\ref{Fig#1} and \ref{Fig#2}}
\newcommand{\listpar}{\setlength{\topsep}{0pt}\setlength{\itemsep}{-0.25\baselineskip}}
\begin{document}
\bibliographystyle{apsrev}
\title{Magnetization steps and cluster-type statistics for a diluted
Heisenberg antiferromagnet on the square lattice:
Models with two exchange constants}
\author{Yaacov Shapira}
\email{yshapira@granite.tufts.edu} \affiliation{Department of
Physics and Astronomy, Tufts University, Medford, MA 02155}
\author{Valdir Bindilatti}
\email{vbindilatti@if.usp.br} \affiliation{Instituto de
F\'{\i}sica, Universidade de S\~{a}o Paulo,\\ Caixa Postal 66.318,
05315--970 S\~{a}o Paulo-SP, Brazil}
\date{\today}
\begin{abstract}
Magnetization steps (MST's) from a strongly diluted antiferromagnet on the
square lattice are discussed theoretically.
Thermal equilibrium, at temperature $T$ and magnetic field $B$,  is assumed.
Two specific cluster models with the largest and second-largest exchange
constants, $J^{(1)}$ and  $J^{(2)}$, respectively, are considered in detail.
In the $J_1$-$J_2$ model, $J^{(1)}$ is the nearest-neighbor (NN) exchange constant
$J_1$, and $J^{(2)}$ is the second-neighbor exchange constant $J_2$.
In the $J_1$-$J_3$ model, $J^{(1)}{=}J_1$, and $J^{(2)}$ is the third-neighbor exchange
constant $J_3$.
For these two  models, all cluster types of sizes $n_c{\le}5$ are identified,
and their statistics is expressed using perimeter polynomials for cluster types.
All cluster types with sizes $n_c{>}1$ give rise to MST's.
Calculated curves of the isothermal magnetization $M$ as a function of $B$ are
given for widely different ratios $J^{(2)}{/}J^{(1)}$.
Some of the information contained in these magnetization curves is conveyed more
directly by the derivative curves, $\rmd M/\rmd B$ versus $B$.
The series of peaks in the derivative curve is called the ``MST spectrum.''
This spectrum is much simpler when the ratio $J^{(2)}{/}J^{(1)}$ is very small.
A detailed discussion of the exchange-bond structure and MST spectra for cluster
models with a very small ratio $J^{(2)}{/}J^{(1)}$ is given in the following
paper.
The ``parent'' models of the $J_1$-$J_2$ model are the $J_1$ and $J_2$ models.
For the $J_1$-$J_3$ model the parent models are the $J_1$ and $J_3$ models.
Relations between the $J_1$-$J_2$ and the $J_1$-$J_3$ models and their respective parent
models are discussed.
The three parent  cluster models ($J_1$-, $J_2$-, and $J_3$-models) are
identical, except for the neighbor  associated with  the one exchange constant
$J$ that is included in the model.
Common properties of such  ``isomorphic'' cluster models are discussed.
\end{abstract}
\maketitle
\section{\label{s:I-intro}INTRODUCTION}
An earlier theoretical paper\cite{Bindi05prb} (hereafter I) on magnetization
steps from a diluted Heisenberg antiferromagnet on the square lattice was
largely devoted to the nearest-neighbor (NN) cluster model.
To obtain magnetization steps (MST's), the NN exchange constant \J1 must
be antiferromagnetic (AF).
The present paper discusses cluster models with two exchange constants:
the largest,  called \JJ1, and the second-largest, \JJ2.
Both exchange constants are assumed to be AF.

Some of the discussion in the present paper applies to any \JJ1-\JJ2 cluster
model, regardless of the neighbors associated with \JJ1 and \JJ2.
However, the main focus is on the \J1-\J2 and \J1-\J3 models.
In these models $J^{(1)}{=}J_1$, and  \JJ2 is either the second-neighbor
exchange constant \J2, or the third-neighbor exchange constant \J3.

As in I, the following assumptions are made:
1) Thermal equilibrium, at temperature $T$ and magnetic field $B$, prevails.
2) All the magnetic ions are identical.
3) The cation sites form a square lattice, and the magnetic ions are randomly
distributed over these sites.
4) Only a fraction $x$ of all cations sites are occupied by magnetic ions.
This fraction is well below the site percolation concentration $x_c$ for the
relevant cluster model.
5) None of the magnetic interactions is anisotropic.
Background material for the present paper may be found in I,
in a recent review,\cite{Shapira02jap} and in earlier
papers.\cite{Vu92,Bindi98prl,Bindi99jap}

This paper is organized as follows.
The principal results are presented in the main text.
Supplementary material is relegated to appendices.
The main text starts with a brief discussion of the crucial role of cluster
``configurations'' in the theory.
All cluster types in the \J1-\J2 and \J1-\J3 models, subject to the restriction
$n_c{\le}5$ on the cluster size $n_c$, are then identified.
The statistics for these cluster types is expressed using perimeter polynomials
(PP's) for cluster types.
As discussed in I, the PP's for cluster types are analogous to the conventional
PP's for cluster sizes.\cite{Stauffer94}

The contribution of any cluster type to the magnetization $M$ is calculated by
combining the results for the energy eigenvalues with those for the cluster
statistics of that cluster type.
The total magnetization $M(T,B)$ is the sum of the contributions from all
cluster types.
In some respects the derivative curve, \dMdB\ versus $B$, is more
informative than the magnetization curve, $M$ versus $B$.
Examples of calculated magnetization and derivative curves at constant $T$  are
given for widely different ratios \rJJ.

The statistics for cluster types is independent of the spin $S$ of the
individual magnetic ions.
However, the energy eigenvalues, and therefore the magnetization curve, depend
on $S$.
Although much of the discussion is for any $S$, all the numerical examples are
for $S{=}5/2$, which is the appropriate value for the $\mathrm{Mn}^{2+}$ and
$\mathrm{Fe}^{3+}$ ions.
Both theses ions are $S$-state ions, and they usually have low crystalline
anisotropy.
Such ions are useful for testing theories in which anisotropy is neglected.

The structures of the magnetization and derivative curves are much simpler when
the ratio \rJJ\ is ``very small.''
Cluster models with such widely different magnitudes of \JJ2 and \JJ1  are
called ``lopsided cluster models.''
In addition to their interesting physics, lopsided models are useful because
they apply to many materials.
Lopsided cluster models are discussed in detail in the following
paper.\cite{Bindi06eprint2}
\section{\label{s:II-config}CLUSTER CONFIGURATIONS}
\subsection{\label{ss:IIA}Configurations}
The calculation of the magnetization $M(T, B)$ in any cluster model requires:
1) the identification of all cluster types $c$ in that model;
2) a calculation of the average magnetic moment $\mu_c(T, B)$  per
realization,\cite{note5} for each cluster type $c$; and
3) an evaluation of the probabilities of finding the various cluster types.

Before carrying out the first and third of these tasks it is necessary to
identify all the ``cluster configurations'' that exist in the particular cluster
model.
These cluster configurations are the fundamental building blocks of the theory,
and also of the computer programs that are used to implement the theory.
Cluster configurations were discussed in Sec.~IIIB of I.
The discussion below brings out some new features.
\subsubsection{\label{sss:IIA1}Cluster configurations}
A spin cluster consists of a finite number of exchange-coupled magnetic ions
(spins) that occupy a set of cation sites.
Spin clusters are considered to have the same configuration if and
only if the sets of cation sites occupied by these clusters
can be obtained from each other by symmetry operations of the space group of
the cation structure.
Each one such set of cation sites is called a ``realization'' of the
configuration.

The symmetry operations of the space group of the cation structure are the only
symmetry operations considered in the present work.
They will be referred to, simply, as ``symmetry operations.''
The symmetry operations that are relevant to the present work are the operations
of the $P4m$ space group of the square lattice, including the lattice
translations.

Realizations of the same configuration have the following important geometrical
property.
Starting from one realization, a rigid object can be constructed by
joining all pairs of cation sites in that realization by straight-line segments.
The straight-line segments may be viewed as ``struts'' that give the object its
rigidity.
Rigid objects constructed in this manner from different realizations of the same
configuration, either have identical shapes or are chiral isomers (mirror
images) of each other.
Because of this geometrical property, configurations of clusters are sometimes
viewed as the geometrical shapes of clusters.

\subsubsection{\label{sss:IIA2} Cluster configurations of one specific cluster
model}
A cluster model is specified by the set of exchange constants (the $J$'s) that
are included in the model.
Any spin cluster in this model consists of spins that are coupled to each other,
but not to other spins, by the $J$'s of the model.
Thus, any two spins in a cluster are connected by at least one continuous path
of exchange bonds associated with this set of $J$'s.
No continuous path of such exchange bonds is allowed to exist between spins in
different clusters.

The restriction on the allowed $J$'s is a restriction on the allowed sets of
cation sites associates with the clusters of the model.
For example, in the \J1-\J2 model, any cation site of a cluster must
have a NN site or a 2nd-neighbor site in the same cluster.
The cluster configurations of the model are all the configurations of the sets
of cation sites that are allowed by the $J$'s of the model.

\subsubsection{\label{sss:IIA3}Identical configurations in different cluster
models}
Sometimes, several cluster models are considered.
It may then happen that the set of $J$'s in one cluster model is only a subset
of all the exchange constants in another cluster model.
Any cluster configuration in the cluster model with the fewer $J$'s is then also
a configuration in the model with the larger set of $J$'s.
For example, the exchange constant \J1 of the NN cluster model is a subset of
the exchange constants in the \J1-\J2 model.
Any cluster configuration in the \J1 model is also a configuration in the
\J1-\J2 model.
The converse is not true; many cluster configurations that exist in the \J1-\J2
model do not exist the \J1 model.

For the same reason, all cluster configurations that exist in the \J1 model
also exist in the \J1-\J3 model.
Figure~2 of I shows an example of the same configuration in different cluster
models.
Figure~2(b) of I shows the configuration in the \J1 model, and Figs.~2(c) and
2(d) show the same configuration in the \J1-\J2 and \J1-\J3 models,
respectively.
\subsection{\label{ss:IIB}From cluster configurations to cluster types}
Once a cluster model is specified, it is necessary to identify the cluster types
that exist in the model.
In the present work this identification was carried out by a series of computer
program.
The first set of programs identified all the (different) configurations that
exist in the specified cluster model.
For each configuration, a realization that has one spin at the origin was
generated.
This realization is called the ``prototype'' of the configuration.
Only configurations with no more than 5 cation sites were considered explicitly
in the present work.

A cluster type $c$ is specified by  a cluster size $n_c$ and by  a bond
list.\cite{Bindi05prb}
To identify the cluster types that exist in the model, the prototypes of all the
(different) configurations were first classified by size, i.e., by the number of
spins in the prototype.
The next step was to generate the bond lists for all prototypes of a given size,
$n_c$.
The final step was to identify all prototypes of the same size $n_c$ that  have
identical bond lists.
Such prototypes are, by definition,\cite{note5} realizations of the same cluster
type, $c$.
In fact, they are the prototypes of all the configurations $r_c$ of cluster type
$c$.

To summarize, the classification of the prototypes of different configurations
by both size and bond list leads to:
1) all cluster types $c$, for each cluster size $n_c$;
2) the bond list for each of these cluster types;
and 3) the prototypes of all the configurations of each cluster type.
\subsection{\label{ss:IIC}Statistics of cluster types}
The goal of the statistics is to find, for each cluster type $c$, the
probability $P_c$ that a randomly-chosen spin is in one of the realizations of
this cluster type.
The main assumption is that the magnetic ions are randomly distributed over the
cation sites.
The procedure for calculating $P_c$ as a function of $x$ was outlined in
Sec.~III C of  I.
The procedure starts from the configurations $r_c$ of cluster type $c$.

Any realization of cluster type $c$ must also be a realization of one of the
configurations $r_c$ of that cluster type.
The probability $P_{r_c}$ that a randomly-chosen spin is in some realization of
the configuration $r_c$ is given by Eq.~(4) of I.
This equation contains two parameters that depend on the configuration:
the lattice-combinatorial parameter $n_{r_c}$, and the perimeter $\nu_{r_c}$.
After these two parameters are evaluated for each of the configurations $r_c$,
the probability $P_c$ is obtained by summing $P_{r_c}$ over all the configurations
$r_c$ of cluster type $c$. This sum is given by Eq.~(5) of  I.

The lattice-combinatorial parameter $n_{r_c}$ is the number of (distinct)
realizations of the configuration $r_c$ that have one spin at the origin.
This $n_{r_c}$ depends only on the configuration.
If the same configuration exists in more than one cluster model, then  the
corresponding lattice-combinatorial parameters are the same in all these models.
The computer program that was used to obtain $n_{r_c}$ was based on the
principle that all realizations of a configuration can be generated from the
prototype of the configuration by applying the symmetry operations of the $P4m$
space group, including the lattice translations.
If the same realization was generated  by different symmetry operations, the
count for $n_{r_c}$  included this realization only once.

In contrast to $n_{r_c}$, the perimeter $\nu_{r_c}$ depends not only on the
configuration but also on the cluster model.
If the same configuration exists in two cluster models, the perimeters  in the
two models are, in general, not equal.
For example, any configuration that is present in the \J1 model is also present
in the \J1-\J2 model.
The perimeter in the \J1 model will be called the \J1 perimeter.
Given any realization of the configuration $r_c$, the \J1 perimeter is the
number of (cation) sites that are NN's of the sites in the realization, but are
not themselves sites of the realization.
The perimeter in the \J1-\J2 model  (called the \J1-\J2 perimeter) is the
number of sites that are either NN's and/or 2nd-neighbors of the sites in the
realization, but are not themselves sites of the realization.
Therefore, for the same configuration, the \J1-\J2 perimeter is larger than the
\J1 perimeter.
As a consequence, the probability that a randomly-chosen spin is in a
realization of this configuration will be lower in the \J1-\J2 model than in
the \J1 model [See Eq.~(4) of I].
Similar results apply to a configuration that exists in both
the \J1 and the \J1-\J3 models.

\section{\label{s:III-types}CLUSTER TYPES }
\begin{figure*}\includegraphics[scale=1]{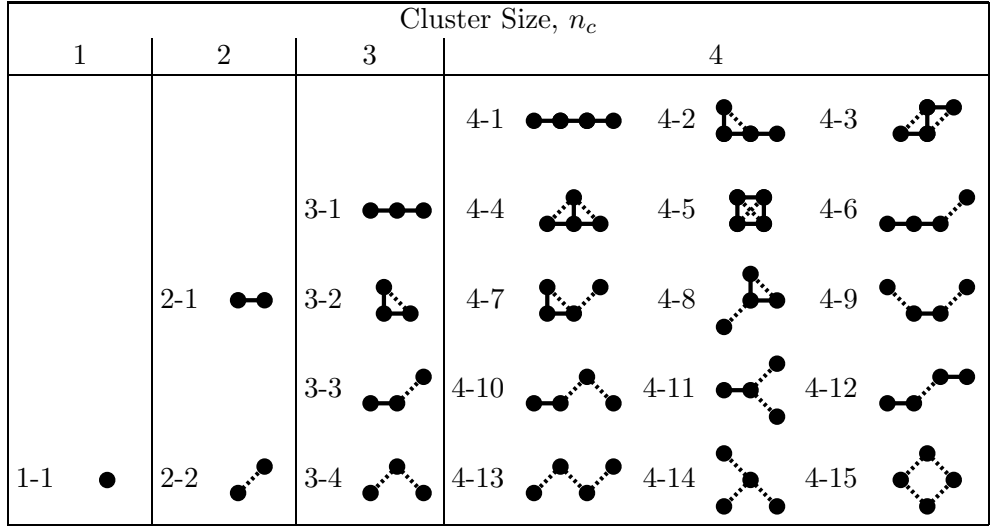}
\caption{\label{Fig1}Cluster types of the \J1-\J2 model, up to cluster size $n_c{=}4$.
Solid circles represent spins. Solid and dotted lines represent \J1 bonds and
\J2 bonds, respectively. The labels for the cluster types are discussed in the text.}
\end{figure*}

\begin{figure*}\includegraphics[scale=1]{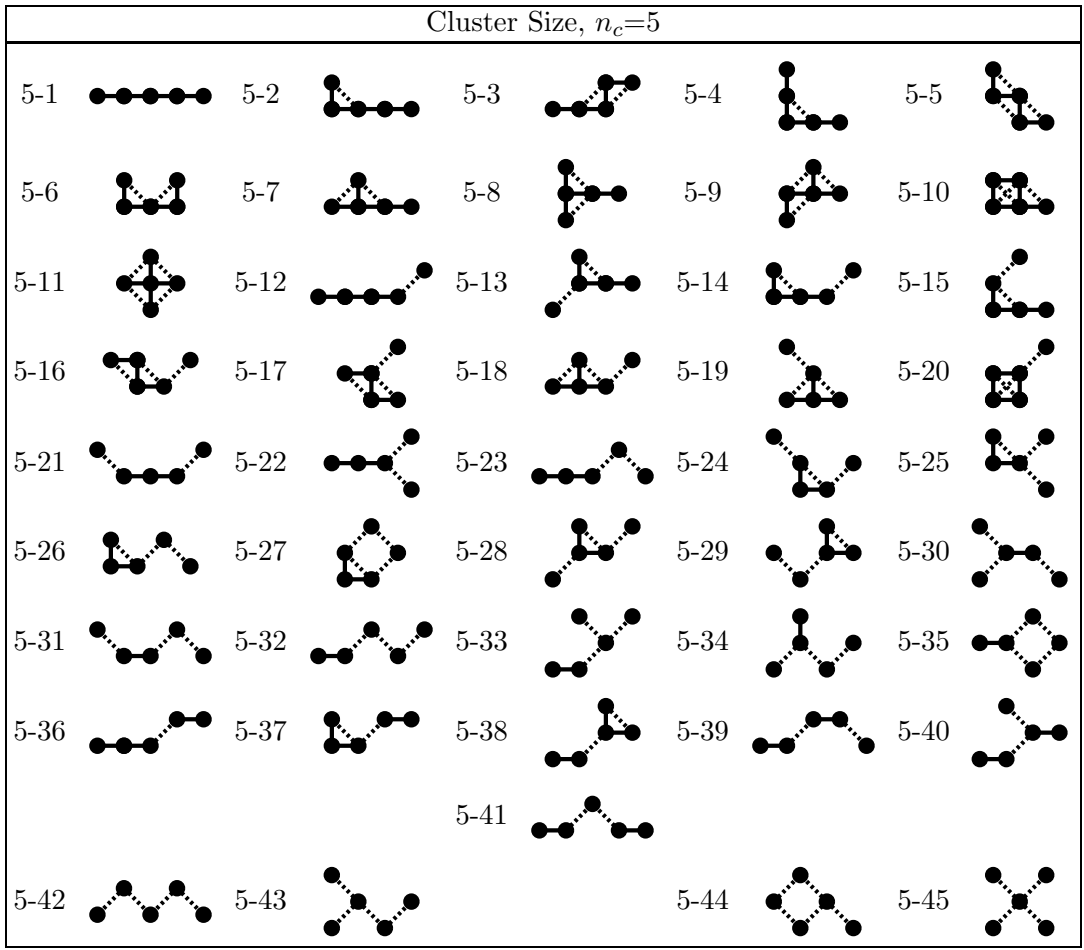}
\caption{\label{Fig2}  Cluster types of quintets $(n_c{=}5$) in the \J1-\J2
model.}
\end{figure*}

\subsection{\label{ss:IIIA}Generic and specific models}
A  \JJ1-\JJ2  cluster model in which the symmetry classes of the neighbors
associated with \JJ1 and \JJ2 are not specified will be called a ``generic'' model.
If the symmetry classes of the neighbors are specified, the cluster model is
``specific'' rather than generic.
In this paper, only two specific \JJ1-\JJ2  models are considered:
the  \J1-\J2 model (with $|J_1|{>}|J_2|$), and the  \J1-\J3 model (with
$|J_1|{>}|J_3|$).
The reason for focusing on these specific models is that exchange constants
often tend to decrease as the distance $r$ of the relevant neighbor increases.
Although the decrease is not always monotonic,\cite{Bindi98prl,Bindi99jap} we
expect that in the vast majority of diluted antiferromagnets with a square cation
lattice, $J^{(1)}{=}J_1$ and \JJ2 is either \J2 or \J3.

The cluster types in the \J1-\J2 model are not trivially related to the cluster
types in the \J1-\J3 model.
That is, the bond lists for the two models cannot be obtained from each other by
replacing the \J2 bonds by \J3 bonds.
The site percolation concentrations for these two specific models are also
different.\cite{Malarz05pre}
The non-trivial dependence of the bond lists, and hence of the cluster types, on
the specific cluster model implies that  bond lists and cluster types cannot be
given for a generic model. They can only be given for a specific model.
\subsection{\label{ss:IIIB}Parent cluster models}
The ``parent'' cluster models of the \J1-\J2 model are the \J1 model and the
\J2 model, each of which has only one of exchange constants of the \J1-\J2
model.
The \J1-\J2 model is not a simple combination of its parent models.
Similarly, the \J1-\J3 model is not a simple combination of the \J1 and \J3
models, which are its parent models.

There are many interesting relations between the \J1-\J2  and  \J1-\J3 models
and their respective parent models.
To avoid repeated interruptions in the main text of the paper, discussions of
these relations are relegated to Appendices.
For the limited purpose of calculating magnetization curves numerically, the
results in these Appendices are not essential.
However, these results give a deeper insight into the physics.
Some of these results will be quoted, and used, in the main texts of the present
paper and of the following paper.
Appendix~\ref{a:iso}, describes the strong similarity (called ``isomorphism'')
between the three parent models, i.e., the \J1, \J2, and \J3 models.
These are the only  ``parent models'' considered in the present work.

\subsection{\label{ss:IIIC} Cluster types in the \J1-\J2 model}
\begin{figure*}\includegraphics[scale=1]{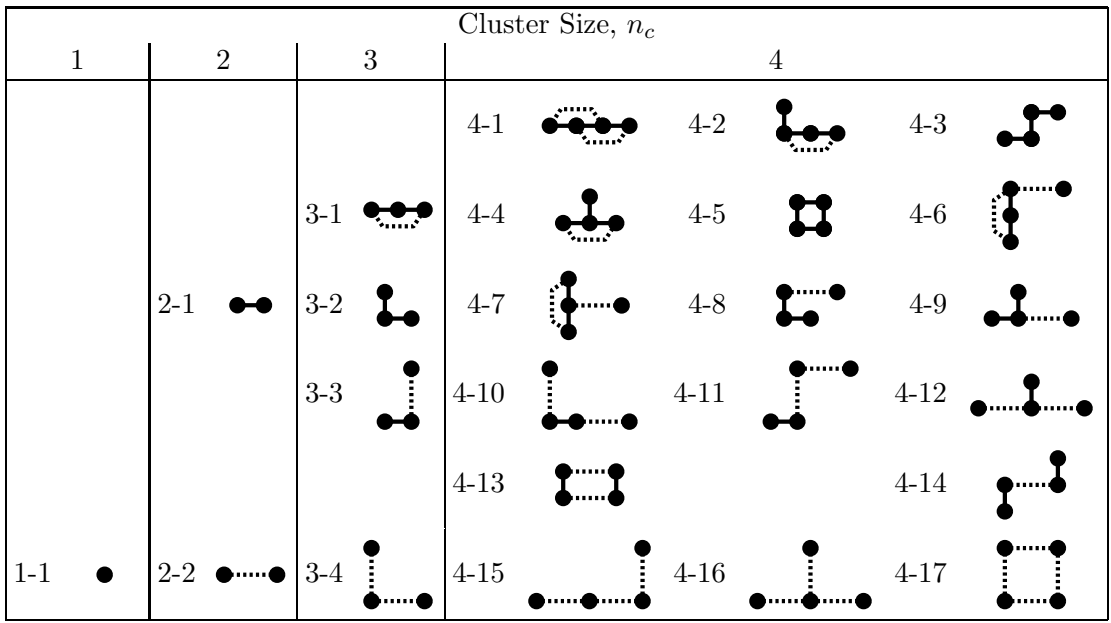}
\caption{\label{Fig3}Cluster types of the \J1-\J3 model, up to cluster size $n_c{=}4$.
The format, including the format of the labels for the cluster types, is similar to that in \Fig1,
except that the dotted lines represent \J3 bonds.}
\end{figure*}

\begin{figure*}\includegraphics[scale=1]{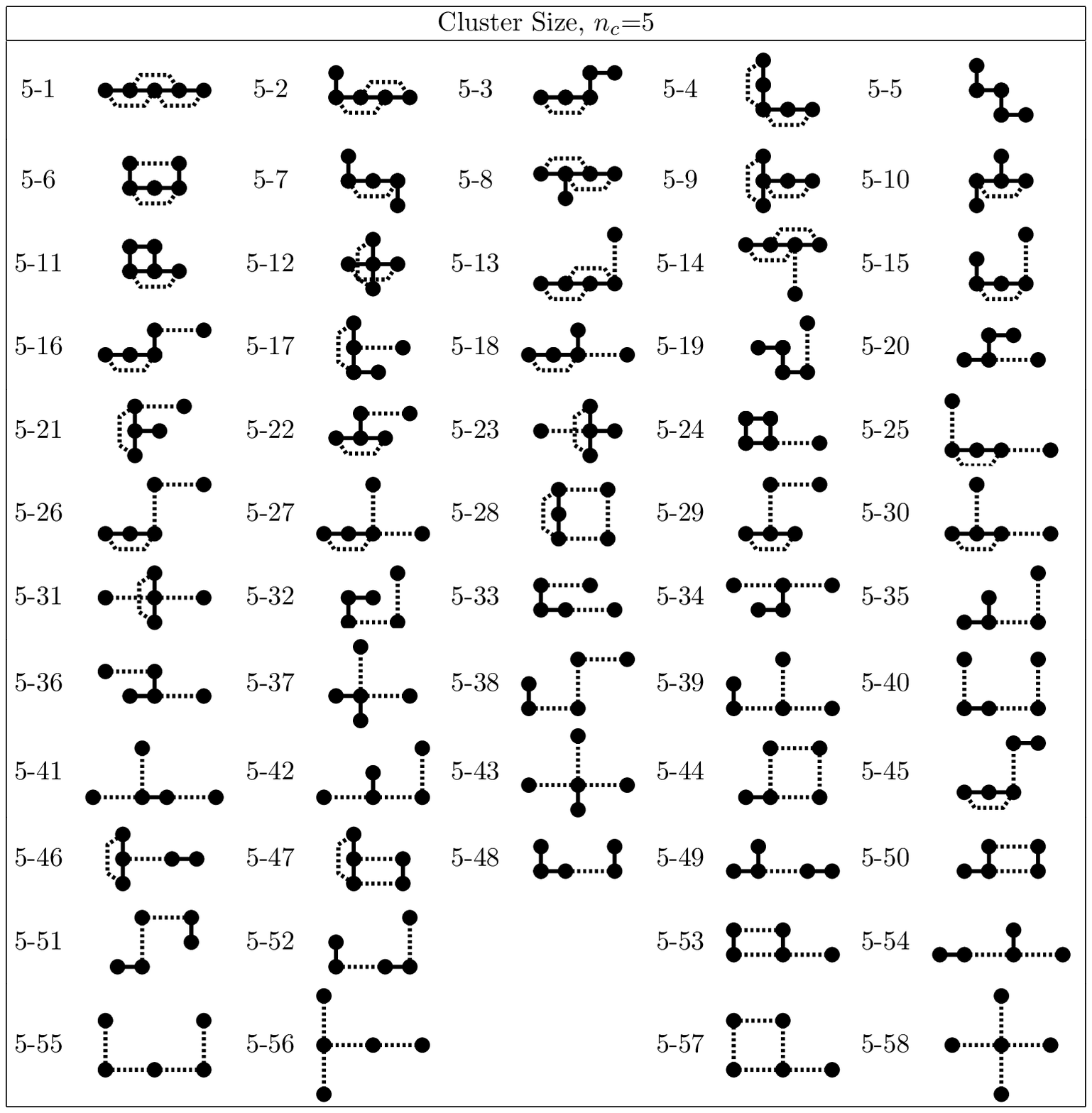}
\caption{\label{Fig4}  Cluster types of quintets $(n_c{=}5$) in the \J1-\J3
model.}
\end{figure*}

\subsubsection{\label{sss:IIIC1}Cluster types}
In the \J1-\J2 model on the square lattice, there are 67 cluster types of sizes
$n_c{\le}5$. These cluster types are shown in \Figs12.
Cluster types of the same size, $n_c$, appear in the same column.
The four columns in \Fig1 show the cluster types with $n_c {=} 1, 2, 3$, and $4$.
The single column in \Fig2 shows the cluster types with $n_c{=}5$.
Spins are represented by solid circles, \J1 bonds by solid lines, and \J2 bonds
by dotted lines.
Only one configuration for each cluster type is shown.
The bond lists for these cluster types, which specify all intra-cluster exchange
interactions,\cite{Bindi05prb} are given in Appendix~\ref{a:bondlists}.

Each cluster type is labeled by two numbers separated by  a hyphen.
The first number is the cluster size $n_c$.
The second is a serial number (SN) within this cluster size.
Thus, the format for any label is [$n_c$-(SN)].
There is only one type of single (type 1-1), but two types of pairs:
2-1 which is a \J1-pair, and 2-2 which is a \J2-pair.
There are four types of triplets ($n_c{=} 3$), 15 types of quartets
($n_c{=}4$), and 45 types of quintets ($n_c{=} 5$).

\subsubsection{\label{sss:IIIC2}Four categories of cluster types}
The number of cluster types in \Figs12 is rather large.
It is therefore useful to classify them.
The classification scheme is not unique. In one scheme the 67 cluster types
in \Figs12 are divided into four broad categories:
\begin{enumerate}\listpar
\item ``single,''  which has no exchange bonds;
\item ``pure \J1'' cluster types, with only \J1 exchange bonds;
\item ``pure \J2'' cluster types,  with only \J2 bonds;
\item ``mixed'' cluster types, with both \J1 and \J2  bonds.
\end{enumerate}

The single (type 1-1) is at the left of the bottom row of \Fig1.
The only pure \J2 cluster types are the other 5 cluster types in the same bottom
row together with the four cluster types in the bottom row of \Fig2.
The only pure \J1 cluster types are: 2-1, 3-1, 4-1, and 5-1.
All the remaining 53 cluster types are ``mixed'' types.

The pure \J1 cluster types, and the pure \J2 cluster types, are related to
cluster types in the (parent) \J1 and \J2 models, respectively.
These relations are discussed in Appendix~\ref{a:relations}.

\subsection{\label{ss:IIID}Cluster types in the \J1-\J3 model}

In the \J1-\J3 model there are 82 cluster types of sizes $n_c{\le}5$.
They are shown in \Figs34.
The format is the same as in \Figs12, except that the dotted lines now
represent \J3 bonds.
The bond lists for the cluster types in \Figs34 are given in
Appendix~\ref{a:bondlists}.

The labels for the  cluster types of the \J1-\J3 model
have the same format as those for the cluster types in the \J1-\J2 model.
Therefore, many of the labels used in the two models models are identical.
When the same label appears in both models, it often refers to different cluster
types (i.e., different bond lists).
Unless it is clear from the context, it is then necessary to specify
the cluster model to which the label refers.

Once again the cluster types in \Figs34 can be divided into four categories:
\begin{enumerate}\listpar
\item  the ``single'' (type 1-1), at the left of the bottom row of \Fig3;
\item  the 9 ``pure \J3'' cluster types, consisting of the other 5 cluster types
in the bottom row of  \Fig3 together with the 4 cluster types in the bottom row of \Fig4;
\item  the 5 ``pure \J1'' cluster types: 2-1, 3-2, 4-3, 4-5, and 5-5;
\item  the 67 remaining cluster types, which are all ``mixed'' types.
\end{enumerate}

\section{\label{s:IV} CLUSTER STATISTICS AND PERIMETER POLYNOMIALS}
\subsection{\label{ss:IVA}Results for small clusters}
The probabilities $P_c$ as a function of $x$ were obtained using the procedure
discussed in Sec.~\ref{ss:IIC}.
Only cluster types with $n_c{\le}5$ were considered.
Some of the results for the \J1-\J2 model are shown \Fig5.
They include the $P_c$'s for the single (type 1-1), for the two types of pairs
(2-1 and 2-2), and for the four types of triplets (3-1, 3-2, 3-3 and 3-4).
These labels for the cluster types refer to \Fig1.
Also shown in \Fig5 are:
the sum $P_4$ of the probabilities for all 15 quartet types in \Fig1;
the sum $P_5$ of the probabilities for all the quintet types in \Fig2;
and the sum $P_{>5}$ of the probabilities for all cluster types with sizes
$n_c{>}5$.
\begin{figure*}\includegraphics[scale=1]{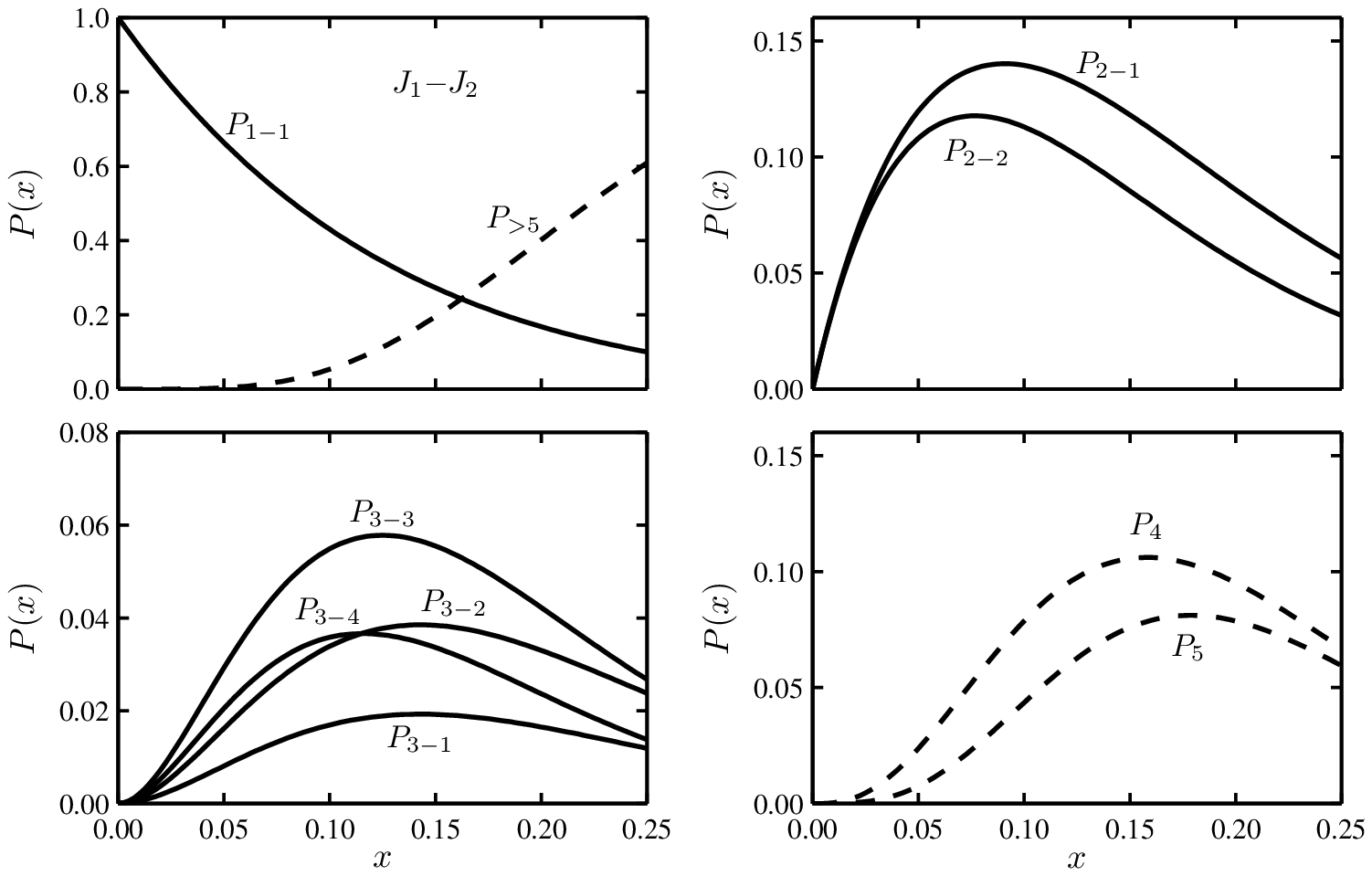}
\caption{\label{Fig5}
Probabilities $P_c$ as a function of $x$ for some cluster types of the
\J1-\J2 model.
The solid curves are for individual cluster types, labeled as in \Fig1.
Curve 1-1 is for singles, curves 2-1 and 2-2 are for the two types of pairs, and
curves 3-1 up to 3-4 are for the four types of triplets.
The dashed curves are probabilities for some combinations of cluster types:
$P_4$ is the sum of the $P_c$'s for all the 15 quartet types in \Fig1;
$P_5$ is the sum of the $P_c$'s for all 45 quintet types in \Fig2;
and $P_{>5}$ is the sum of the probabilities for all cluster types of sizes
$n_c{>}5$.
}
\end{figure*}

\Figure6 shows the corresponding probabilities for the \J1-\J3 model.
In this case the cluster types refer to \Figs34.
In both \Fig5 and \Fig6  the highest value of $x$ is below the relevant site
percolation concentration, $x_c{=}0.407$ for the \J1-\J2 model,
and $x_c{=}0.337$ for the \J1-\J3 model.\cite{Malarz05pre,Peters79}
For the \J1 model, $x_c{=}0.593$.
\begin{figure*}\includegraphics[scale=1]{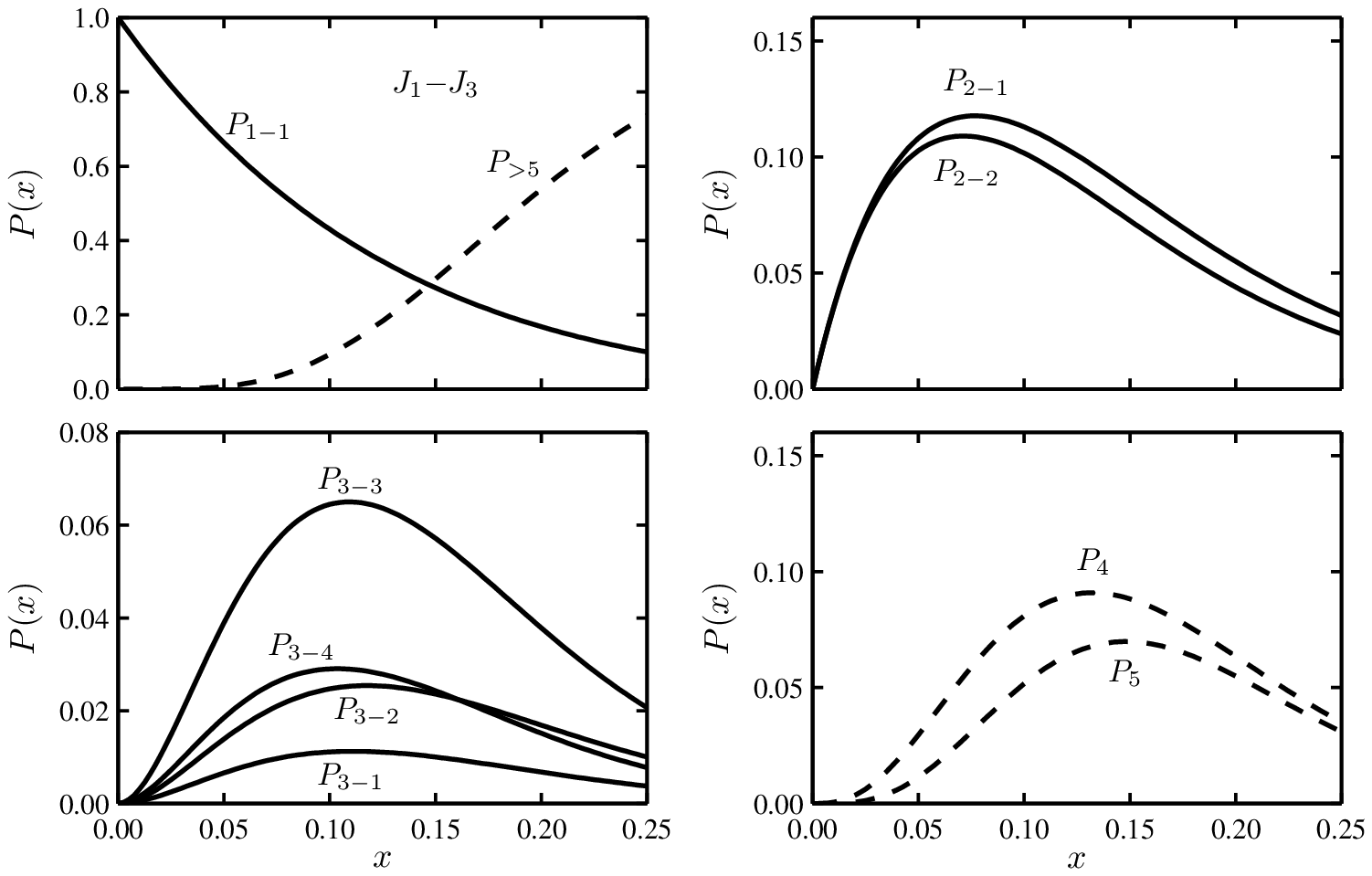}
\caption{\label{Fig6}
Probabilities $P_c$ as a function of $x$ for some cluster types of the \J1-\J3
model.
The labels for the various curves are similar to those in \Fig5, except that
the cluster types refer to \Figs34 for the \J1-\J3 model.}
\end{figure*}

\subsection{\label{ss:IVB}Perimeter polynomials}
As discussed in I, the probability $P_c$ can be expressed succinctly in the form
\begin{equation}
  P_c = n_cx^{n_c{-}1}D_c(q),                    \label{Eq:1}
\end{equation}
where $D_c(q)$ is a polynomial in $q{=}1{-}x$, defined as the perimeter
polynomial (PP) for cluster type $c$.
Appendix~\ref{a:bondlists} gives the PP's for cluster types with sizes
$n_c{\le}5$ in the \J1-\J2 model. These are the cluster types in \Figs12.
Appendix~\ref{a:bondlists} also gives the PP's for cluster types of the \J1-\J3
model  whose sizes are $n_c{\le}5$. These are the cluster types in \Figs34.

For the \J1-\J2 model, the following check on the results for $D_c(q)$ in
Appendix~\ref{a:bondlists} was performed.
The conventional  PP's, $D(q)$, for cluster sizes (not types) were given
for this model by Peters et al.\cite{Peters79}
As expected, the polynomial $D(q)$ for any cluster size $s$ is equal to the sum
$\sum D_c(q)$ of the PP's in Appendix~\ref{a:bondlists} over all cluster types
$c$ that have the size $n_c{=}s$.

Appendix~\ref{a:comments} discusses some relations between the probabilities
$P_c$ for the pure-\J1 and pure-\J2 cluster types in the \J1-\J2 model and the
probabilities for the same cluster types in the (parent) \J1 and \J2 models.
The relations between the probabilities $P_c$ for pure-\J1 and pure-\J3 cluster
types in the \J1-\J3 model and the probabilities for the same cluster types in
the (parent) \J1 and \J3 models are also discussed.

\section{\label{s:V}THE MAGNETIZATION CURVE}
\subsection{\label{ss:VA}Calculation procedure}
The procedure for calculating the magnetization $M(T,B)$ was discussed
previously.\cite{Bindi05prb,Shapira02jap}
Briefly, the Hamiltonian of one realization\cite{note5} of each cluster type $c$
is diagonalized, and the results are used to obtain the average magnetic moment
$\mu_c(T,B)$ per realization.
For singles, $\mu_c(T,B)$ follows the Brillouin function for spin $S$.
For each of the other cluster types, $\mu_c(T,B)$ exhibits a series of MST's as
a function of $B$ at very low $T$.

The total magnetization $M(T,B)$ is a statistically-weighted sum of $\mu_c(T,B)$.
This sum is given by Eq.~(13) of I, namely,
\begin{equation}
 M(T,B) = \sum_c N_c \mu_c(T,B) = N_\mathrm{total}\sum_c
\frac{P_c}{n_c}\mu_c(T,B),
\label{Eq:2}
\end{equation}
where $N_c{=}P_c N_\mathrm{total}/n_c$ is the population of cluster type $c$,
and $N_\mathrm{total}$ is the total number of spins.
The quantities $M$, $N_c$  and $N_\mathrm{total}$ are all either per unit mass
or per unit volume.

The infinite sum in Eq.~(\ref{Eq:2}) cannot be evaluated exactly because
$\mu_c(T,B)$ and $P_c$ (or $N_c$) are known only for a finite number of cluster
types $c$.
Usually they are known only for cluster types whose sizes $n_c$ are no
larger than some maximum size \nmax.
The sum in Eq.(~\ref{Eq:2}) is therefore truncated after the finite sum over
cluster types with $n_c{\le} n_\mathrm{max}$ is evaluated exactly.
The remainder (REM) from clusters of sizes $n_c{>}n_\mathrm{max}$, is then
approximated by the ``remainder correction''  $R(T,B)$.
In the present work, $n_\mathrm{max}{=} 5$, so that the REM is from clusters larger
than quintets.
The remainder correction $R(T,B)$  was obtained by the corrective quintets
(CQUIN's) method.
The application of this method to the NN cluster model was discussed in I.
For a model with two exchange constants the CQUIN's method is considerably more
involved.

The number of spins in the REM is $P_{>5}N_\mathrm{total}$.
The accuracy of the CQUIN's method is not an important issue if $P_{>5}{\ll}1$.
Because $P_{>5}$ increases with $x$, the accuracy is not a significant issue if
$x$ is sufficiently small.
All magnetization curves in the present paper are for $x {\leq} 0.09$,
and are based on the \J1-\J2 model.
Under these conditions, $P_{>5} {\leq}  3.6\%$, so that errors in $M$ resulting from
the CQUIN's method are not significant.
The description of the CQUIN's method is postponed to the following paper, which
also includes some examples for higher $x$.

\subsection{\label{ss:VB}The two reduced magnetic fields}
The two exchange constants, \JJ1 and \JJ2, lead to two energy scales for the
Zeeman energy.
In analogy to Eq.~(10) of I, the primary reduced magnetic field $b_1$ is defined
as
\begin{subequations}\label{Eq:3}
\begin{equation}
  b_1 = g\muB B/|J^{(1)}|.          \label{Eq:3a}
\end{equation}
The secondary reduced magnetic field $b_2$ is
\begin{equation}
  b_2 = g\muB B/|J^{(2)}|.          \label{Eq:3b}
\end{equation}
\end{subequations}
Thus, at any given $B$,
\begin{equation}
       {b_1}/{b_2} = \left|{J^{(2)}}/{J^{(1)}}\right|  < 1.\label{Eq:4}
\end{equation}
The reduced magnetization $m$ is defined as in I. That is,
\begin{equation}
                m = M/M_0,                   \label{Eq:5}
\end{equation}
where $M_0$   is the true  saturation value of $M$.
The reduced parameters $b_1$, $b_2$, and $m$,  will be used  in plots and
discussions of the magnetization curves.

\subsection{\label{ss:VC}MST's from the two different types of pairs}
At a low $T$ the calculated magnetization curve, $M$ versus $B$, includes a
superposition of many series of MST's.
Each series arises from some cluster type $c$ of size
$2{\le}n_c{\le}n_\mathrm{max}$.
The magnitude $(\Delta M)_c$ of a magnetization jump at each MST in the series
is proportional to the population $N_c$ of  the relevant cluster type.
For low $x$ the largest jumps $(\Delta M)_c$ are for \JJ1 pairs and \JJ2 pairs.
In both \J1-\J2 and \J1-\J3  models these pairs are cluster types 2-1 and 2-2,
respectively (see \Figs13).

The MST's from \JJ1 pairs occur at the primary reduced fields
\begin{subequations}\label{Eq:6}
\begin{equation}
b_1 = 2, 4, 6,\dots,4S.             \label{Eq:6a}
\end{equation}
The MST's from  \JJ2 pairs occur when the secondary reduced field has the same
values, i.e., at
\begin{equation}
b_2 = 2, 4, 6,\dots,4S.             \label{Eq:6b}
\end{equation}
\end{subequations}
Experimental values of the magnetic fields $B$ at the MST's from pairs are often
used to determine \JJ1 and \JJ2.
The temperature requirement for resolving the MST's from \JJ2 pairs is
$\kB T{<}|J^{(2)}|$.
This is a more stringent requirement than $\kB T{<}| J^{(1)} |$ for resolving
the MST's from \JJ1 pairs.

Equations (\ref{Eq:6a}) and (\ref{Eq:6b}) use the reduced fields $b_1$ and $b_2$.
Using the magnetic field $B$, instead of the reduced fields,
the ranges of  $B$ for the MST series from the two types of pairs may or may not overlap.
The condition for avoiding overlap is
\begin{equation}
            {J^{(2)}}/{J^{(1)}} < {1}/{2S}.               \label{Eq:7}
\end{equation}
For magnetic ions with $S{=}5/2$ (e.g., $\mathrm{Mn^{2+}}$ or $\mathrm{Fe^{3+}}$)
overlap is avoided if $J^{(2)}{/}J^{(1)}{<}0.2$.

\subsection{\label{ss:VD}Two examples of magnetization curves for very low $x$}
The main purpose of the following two examples is to illustrate the dependence
of the MST pattern on the ratio \rJJ.
The examples are for $x{=}0.01$ in the \J1-\J2 cluster model.
To optimize the resolution of the spectra, the examples are for $T{=}0$.
The value $S{=}5/2$ is assumed.

\Figures78 are for $J_2{/}J_1{=}0.28$.
\Figure7 shows the reduced magnetization $m$ as a function of the primary
reduced magnetic field $b_1$.
MST's from \J1 pairs (cluster type 2-1) are indicated by long arrows, and
those from \J2 pairs (cluster type 2-2) by shorter arrows.
The main part of \Fig7 shows the magnetization curve up to $b_1=7.5$.
The full magnetization curve is shown in the inset.
\begin{figure}\includegraphics[scale=1]{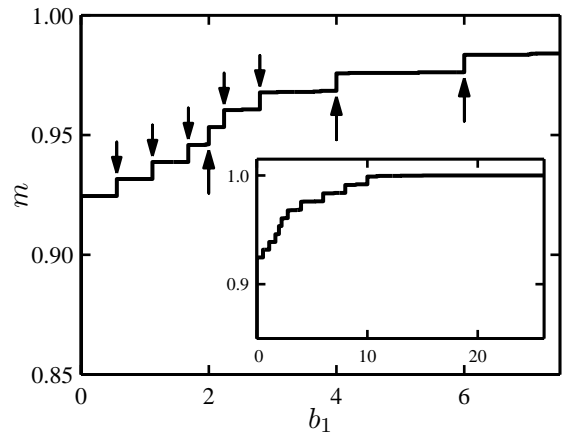}
\caption{\label{Fig7}Magnetization curve at $T{=}0$, calculated from the \J1-\J2
model using the  parameters $x{=}0.01$, $S{=}5/2$, and $J_2{/}J_1{=}0.28$.
The ordinate is the reduced magnetization, $m{=}M{/}M_0$, where $M_0$ is the true
saturation magnetization. The abscissa is the primary reduced magnetic field
$b_1{=}g\muB B{/}|J_1|$, up to 7.5.
MST's from \J1 pairs  (cluster type 2-1) are indicated by long upward arrows.
MST's from \J2 pairs  (cluster type 2-2) are indicated by shorter downward
arrows. The inset shows the full magnetization curve, up to complete
saturation.}
\end{figure}
\begin{figure}\includegraphics[scale=1]{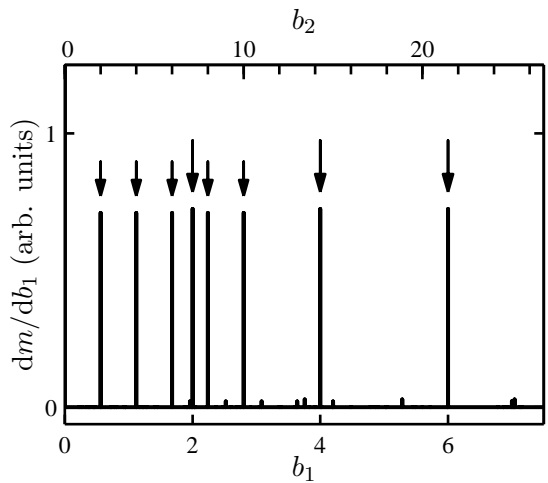}
\caption{\label{Fig8}The derivative of the magnetization curve in \Fig7.
The ordinate is the derivative  $\rmd m/\rmd b_1{=}(|J_1|{/}g\muB M_0) (\rmd M{/}\rmd B)$.
The lower abscissa scale is for the primary reduced field $b_1$.
The upper abscissa scale is for the secondary reduced magnetic field $b_2$.
The derivative peaks at MST's from cluster types 2-1 (\J1 pairs) and 2-2 (\J2
pairs) are indicated by long and short arrows, respectively.}
\end{figure}

\Figure8 shows the derivative \dMdB, in normalized units,\cite{note10}
in fields up to $b_1=7$, corresponding to the main part of \Fig7.
The upper abscissa scale is for the secondary reduced magnetic field $b_2$.
The derivative peaks at MST's from \J1-pairs (type 2-1) and from \J2-pairs (type 2-2),
are indicated by long and short arrows, respectively.
Their reduced magnetic fields  are given by Eqs.~(\ref{Eq:6a}) and (\ref{Eq:6b}).
Because the ratio $J_2{/}J_1{=}0.28$ is higher than 0.2, the field ranges for the
series of MST's from the two types of pairs overlap.

\Figure9 shows the magnetization (top) and derivative (bottom) curves for
$x{=}0.01$ when the ratio $J_2{/}J_1{=}0.028$, i.e., smaller by a factor of
10 compared to the ratio in \Figs78. All other parameters are the same as for
\Figs78.
Because the ratio $J_2{/}J_1$ is now well below 0.2, the MST series from \J1- and
\J2-pairs occur in field ranges that do not overlap.
In the ``gap'' between these two field ranges, the magnetization (upper curve)
exhibits a plateau of apparent saturation, labeled as $m_s$.
The apparent saturation value, $m_s{=}0.961$, agrees with the value of $m_s$
obtained from the NN cluster model for this $x$ (Ref.\onlinecite{Bindi05prb}).
However, in contrast to the NN cluster model, the apparent saturation is reached
only after the series of MST's from the \J2 pairs (type 2-2) is completed.
Clearly, the MST pattern  for $J_2{/}J_1{=}0.028$ (\Fig9) is much simpler than
that for $J_2{/}J_1{=}0.28$ (\Figs78).

\begin{figure}\includegraphics[scale=1]{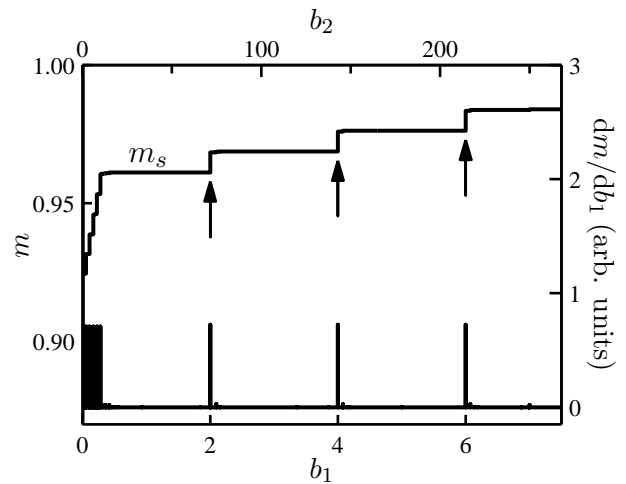}
\caption{\label{Fig9}Zero-temperature magnetization curve (upper curve) and its
derivative (lower curve) for $x{=}0.01$.
These curves are calculated using the \J1-\J2 model and the ratio
$J_2{/}J_1{=}0.028$.
As in \Fig8 the lower and upper abscissa scales are for $b_1$ and $b_2$,
respectively.
The left ordinate scale is for $m$. The right ordinate scale is for \dmdb.
The arrows indicate the locations of the MST's from \J1 pairs (type 2-1).
The MST's from \J2 pairs (type 2-2) are bunched up at low fields.
The plateau of apparent saturation is labeled as $m_s$.}
\end{figure}

\Figure7 for $J_2{/}J_1{=}0.28$  shows a short magnetization plateau
immediately after the initial magnetization rise.
This plateau, which ends at the first MST from \J2 pairs (not \J1 pairs), is not
the plateau of apparent saturation predicted by the NN cluster model.
The value $m{=}0.925$ at the first short plateau in \Fig7 is well below the
apparent saturation value $m_s{=}0.961$ in the NN cluster model.
The reason for the lower value is that the series of MST's from \J2 pairs has
not been completed before the start of this short plateau.

\section{\label{s:VI}THE MST SPECTRUM}
\subsection{\label{ss:VIA}Spectrum}
The derivative \dMdB\ as a function of $B$ exhibits a peak at each MST.
The pattern of the peaks in the derivative curve will be called the ``MST spectrum.''
\Figure8 and the lower curve in \Fig9 are examples of such MST spectra.
Each peak in  \dMdB\  is a ``spectral line.''
Two or more spectral lines associated with MST's arising from different cluster
types may overlap.
When a spectral line is due to only one MST from one cluster type $c$, the
integral of the spectral line with respect to $B$ is equal to the magnetization
jump $(\Delta M)_c$.

There are several advantages of using the spectrum.
A plot of the spectrum, \dMdB\ versus $B$, is very effective in conveying
information visually.
Another advantage is that the calculated spectrum is the \emph{exact} spectrum
from cluster types with $n_c{\le}n_\mathrm{max}$.

As discussed earlier, the infinite sum in Eq.~(\ref{Eq:2}) is split into a sum
over clusters with $n_c{\le}n_\mathrm{max}$, and a remainder (REM) from larger
clusters ($n_c{>}5$ in the present work).
The reason for the split is that the Hamiltonians of clusters with
$n_c{>}n_\mathrm{max}$ have not been diagonalized.
The finite sum is evaluated exactly, but the REM is approximated by $R(T,B)$.
The derivative of  $R(T,B)$ does not give the spectral lines from the clusters
with  $n_c{>}n_\mathrm{max}$.
These lines can be obtained only if cluster Hamiltonians for
$n_c{>}n_\mathrm{max}$ are diagonalized.

A wealth of information, such as accurate exchange constants, can be obtained
from a comparison of an experimental spectrum with the calculated exact spectrum for
$n_c{\le}n_\mathrm{max}$.
All the spectra shown in this paper are the derivatives of the exact finite sum
up to $n_\mathrm{max}{=5}$, without the remainder (see Ref.~\onlinecite{note11}).

\subsection{\label{ss:VIB}Additional examples of MST patterns and spectra}
The examples in Sec.~\ref{ss:VD}, for $J_2{/}J_1{=}0.28$ and $0.028$, assumed
$x{=}0.01$.
The following examples are for the same ratios \rJ2, but for higher $x$.
All other parameters ($S{=}5/2$, $T{=}0$), and the cluster model  (\J1-\J2), are
the same as in Sec.~\ref{ss:VD}.
An example of a zero-temperature spectrum calculated from the \J1-\J3 model will
be given in the following paper.\cite{Bindi06eprint2}
Calculated spectra at finite temperatures will be shown in
Ref.~\onlinecite{exp15mK} in connection with analysis of experimental data.

\subsubsection{\label{sss:VIB1}  Magnetization curves and spectra for $x{=}0.09$}
Magnetization curves and spectra for $x{=}0.09$ are shown in \Figs{10}{11}.
From \Fig5 the probabilities $P_c$ for all four triplet types increase when $x$
changes  from $0.01$ to $0.09$.
The largest increase is for triplet type 3-3, which is a \J1 pair attached to
a third spin by a \J2 bond.
As a result, spectral lines from type 3-3 triplets are readily seen in
\Figs{10}{11}.

\Figure{10} shows the MST pattern (top curve) and the MST spectrum (lower curve)
for $x{=}0.09$ when $J_2{/}J_1{=}0.28$.
The range of the primary reduced field is $b_1{<}7.5$.
The cluster types responsible for some prominent spectral lines are indicated.
Clearly, for $J_2{/}J_1=0.28$ the spectrum is quite complicated
because of the overlap between different series of MST's from different cluster
types.
\begin{figure}\includegraphics[scale=1]{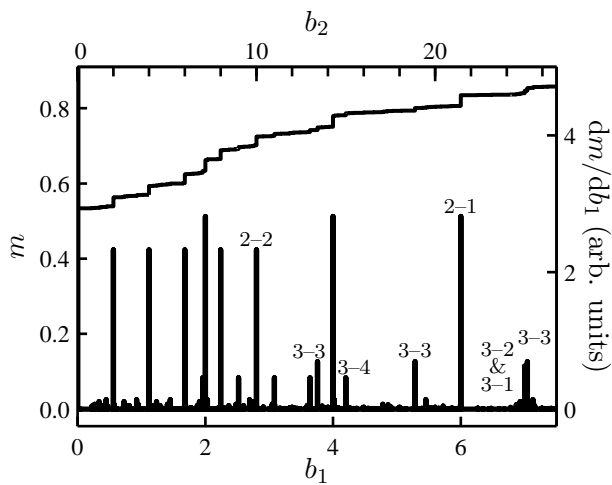}
\caption{\label{Fig10}The reduced magnetization $m$ (top curve) and the MST
spectrum (lower curve) at  $T{=}0$ for $x{=}0.09$, $S{=}5/2$.
These curves are for $J_2{/}J_1{=}0.28$.
Only the results in  the range $b_1{<}7.5$ are shown.
The cluster types responsible for some of the spectral lines are indicated.}
\end{figure}

\begin{figure}\includegraphics[scale=1]{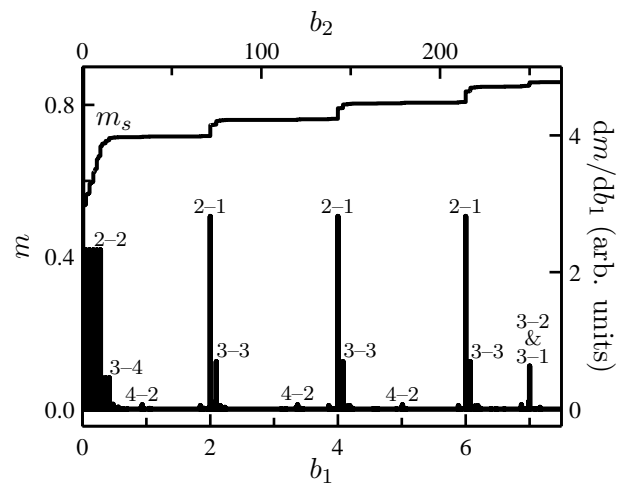}
\caption{\label{Fig11}Zero-temperature magnetization curve (top curve) and
spectrum (lower  curve) for  $x{=}0.09$, calculated from the \J1-\J2  model when
$J_2{/}J_1{=}0.028$.
The range of the primary reduce field is limited to $b_1{<}7.5$.
The plateau of apparent saturation is labeled as $m_s$.}
\end{figure}

\Figure{11} shows the results for the same $x$ when $J_2{/}J_1{=}0.028$.
For this much lower \rJ2 ratio, the spectrum is much simpler.
All the discernable spectral lines occur in two separate field ranges.
The top of the low-field range is slightly above $b_1{=}0.42$ where the series
from type 3-4 triplets ends.
The bottom of the high-field range is near $b_1{=}0.935$ where a small MST from
quartet type 4-2 is barely discernable.
The two field ranges are separated by a gap, i.e., by a field range in which
there are no discernable spectral lines.
The absence of discernable lines implies that $m$ has reached a plateau.
The value $m{=}0.715$ at this plateau agrees with the apparent saturation value
$m_s$ calculated from the NN cluster model.

In the field range of \Fig{11} the \J1 pairs, which exist both in the \J1
and the \J1-\J2  models, give rise to large MST's at $b_1{=}2, 4, 6$.
These are the 2-1 1ines in \Fig{11}.
Near each 2-1 line there is also a line from the 3-3 triplets.
The 3-3 lines do not exist in the \J1 model.
Each 3-3 line together with the stronger nearby 2-1 line may be viewed as a
fine structure (FS) that has evolved from a single spectral line,
due to \J1 pairs, in the \J1 model.
The separation $\Delta b_2$, in the secondary reduced field, between the 3-3
line and the nearby 2-1 line is of order $1$.
The corresponding magnetic field separation $\Delta B$ is $g\muB \Delta B{\sim}|J_2|$.

\Figure{11} also shows a spectral line at $b_1{=}7$ labeled as 3-2 \& 3-1.
It corresponds to the coincidence of the first MST from triplets of type 3-1
and the first MST from triplets of type 3-2.
The ``intensity'' of this combined line is just the sum of the two  intensities.
The other lines from the 3-1 and 3-2 triplets, at $b_1{=}9,11,13$ and $15$
(above the field range of \Fig{11}), also coincide when $J_2{/}J_1{=}0.028$.
\Figure10 shows that the 3-1 and 3-2 lines still coincide when $J_2{/}J_1{=}0.28$.
It can be shown that this remains true as long as $J_2{/}J_1 {\le} 1$.
For additional details see Ref.~\onlinecite{note:coincidence}.

\subsubsection{\label{sss:VIB2}Spectrum for $x{=}0.20$}
For $x{=}0.20$,  approximately 40\% of the spins are in clusters of sizes
$n_c{\ge}5$ (see \Fig5).
When such a large fraction of the spins are in the  REM of Eq.~(\ref{Eq:2}),
the accuracy of the CQUIN's method  of treating the magnetization from the REM
is open to question.
Under these circumstances, a plot of the exact spectrum for clusters with
$n_c{\le}5$ is probably preferable to a plot of the magnetization curve.

\begin{figure}\includegraphics[scale=1]{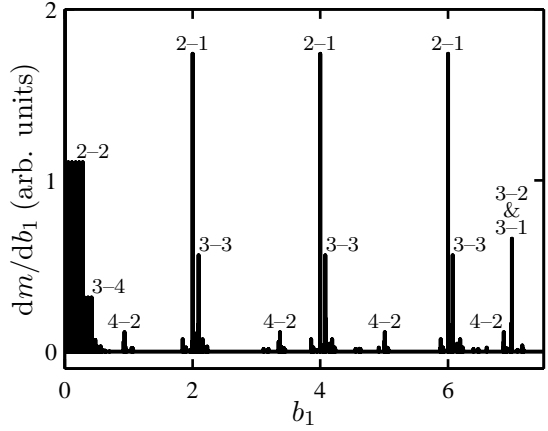}
\caption{\label{Fig12}Zero-temperature spectrum in the \J1-\J2  model  for
$J_2{/}J_1{=}0.028$ when $x{=}0.20$.
Cluster types responsible for some spectral lines are indicated.}
\end{figure}

\begin{figure}\includegraphics[scale=1]{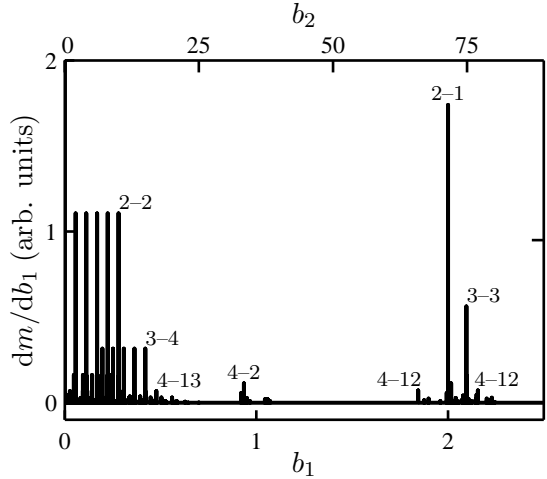}
\caption{\label{Fig13}Expanded view of the spectrum in \Fig{12} $(x{=}0.20)$ for
the range $0{<} b_1 {<} 2.5$.}
\end{figure}

The calculated spectrum for $x{=}0.20$ will be shown only for the low ratio
$J_2{/}J_1{=}0.028$.
\Figure{12} shows this spectrum in the range $0{<}b_1{<}7.5$, at $T{=}0$.
Cluster types responsible for some of the spectral lines are indicated.
An expanded view of the spectrum in the range $b_1{<}2.5$ is shown in \Fig{13}.
The comparison of \Figs{12}{13} with  \Fig{11} shows that the
increase of $x$ from $0.09$ to $0.20$ has led to the following changes:
\begin{enumerate}\listpar
\item  There are new discernable lines. Some of these new lines are in the FS
near $b_1{=}2$.
Two of the lines in this FS are from cluster type 4-12.
One of these 4-12 lines is  slightly above the 2-1 line
(from pure \J1 pairs in the \J1-\J2  model), and and the other 4-12 line is
slightly below it .
Actually there is even a stronger new line in the same FS from quartets of type
4-10. However, this line is not resolved because it is  very close to the still
stronger line from the 3-3 triplets.
\item  A gap, in which there are no discernable spectral lines, separates the
low-field and high-field parts of the spectrum.
On the scale of  \Fig{13} there are no discernable lines between $b_1{=}0.63$
and $b_1{=}0.92$.
\end{enumerate}

\section{\label{s:VII}LOPSIDED MODELS}
The spectra for a ``very small'' ratio \rJ2, are shown in Figs.~\ref{Fig9},
and \ref{Fig11}--\ref{Fig13}.
These spectra are relatively simple.
They suggest\cite{note7A} the following five features for such small \rJ2 ratios.
\begin{enumerate}\listpar
\item The spectrum consists of a low-field part and a high-field part,
separated by a gap in which there are no discernable spectral lines.
\item In the field range of the gap, the magnetization exhibits apparent
saturation, with an apparent saturation value $m_s$  equal to that
given by the NN cluster model.
\item In the high-field part of the spectrum, many spectral lines that exist
in the parent \J1 model develop a FS.
The most conspicuous FS evolves from those lines in the \J1 model
that are due to \J1-pairs, i.e., the lines at $b_1{=}2, 4, 6,$ etc.
\item Separations between adjacent lines in the FS that has evolved from a
single line in the \J1 model are of order $\Delta b_2 {\sim} 1$.
The corresponding separations $\Delta B$ are of order $g\muB B\Delta B{\sim}|J_2|$.
\item In the low-field part of the spectrum, the separations between adjacent
lines are also of order $\Delta b_2{\sim}~1$, or $g\muB B\Delta B{\sim}|J_2|$.
\end{enumerate}
The same five features are also found in simulations that use the \J1-\J3 model
when the ratio \rJ3 is ``very small.''

When the ratio \rJJ\ is sufficiently small that the five features
listed above appear in the spectrum, the \JJ2-\JJ1  model is called
``lopsided.''
Spectra from lopsided models are discussed in detail in the following paper.
Far from a mere curiosity, lopsided models actually apply to many materials.
In Ref.~\onlinecite{exp15mK}, which appears in this issue, the theoretical
results for lopsided models will be used to interpret data
obtained in $\mathrm{(C_3NH_3)_2Mn_xCd_{1-x}Cl_4}$  near $20\;\mathrm{mK}$.
\acknowledgments
This work was supported by CNPQ and FAPESP. Travel funds for Y.~S. were
also provided by FAPESP.

\appendix
\section{\label{a:iso}Isomorphism  of the parent \J1, \J2, and \J3 models}
Starting from Fig.~1 of I, and using symmetry arguments\cite{note12} one can
show that the cluster types of the \J1-model, of the \J2-model, and of the
\J3-model, are identical except for the difference in the symmetry class
of the neighbor associated with the (only one) exchange constant $J$
that is included in the model.
That is, the bond lists for all cluster types in the \J2 model can be
obtained from those in the \J1 model by replacing the label 1 for NN's
by the label 2 for 2nd neighbors.
The bond lists in the \J3 model are also the same, except that the label 3
for the 3rd neighbors replaces the label 1.
Different cluster models, each with only one exchange constant, whose bond
lists are related in this manner will be called ``isomorphic.''
Corresponding cluster types in isomorphic cluster models will be called
isomorphic cluster types.

The exchange part of the cluster Hamiltonian is specified by the bond list for
that cluster type.\cite{Bindi05prb}
Cluster Hamiltonians of isomorphic cluster types are identical except for the
numerical value of the  only $J$ (see Ref.~\onlinecite{note7}).
Each cluster type, except the single, leads to a series of MST's.
The magnetic fields at the MST's originating from isomorphic cluster types are
the same except for a scale factor which is proportional to the $J$ in the model.

For the three parent cluster models considered here, the probabilities $P_c$ for
isomorphic cluster types are the same.
The site percolation concentrations for these three models are also the
same.\cite{Malarz05pre}
It is noteworthy that for the square lattice the \J4-model is not isomorphic to
the \J1, \J2, and \J3 models.
This difference  can be understood from Fig.~1 of I. The number of 4th neighbors
surrounding the central cation site is 8, compared to 4 for the 1st, 2nd, and 3rd neighbors.

\section{\label{a:bondlists}Bond lists and perimeter polynomials for the \J1-\J2
and the \J1-\J3 cluster models}
Tables~\ref{t:I} and \ref{t:II} in give the bond lists, and the perimeter
Polynomials, $D_c(q)$, for the cluster types of the \J1-\J2 model.
The cluster types $c$, limited to sizes $n_c{\leq}5$, are those shown in \Figs12.
\begin{table}
\renewcommand{\arraystretch}{1.25}
\begin{ruledtabular}
\begin{tabular}{ccc}
Cluster type, $c$&Bond List&$D_c(q)$\\\hline
1-1&\{\}&$q^{8}$\\
\hline
2-1&\{1\}&$2q^{10}$\\
2-2&\{2\}&$2q^{12}$\\
\hline
3-1&\{11;0\}&$2q^{12}$\\
3-2&\{11;2\}&$4q^{12}$\\
3-3&\{12;0\}&$8q^{14}$\\
3-4&\{22;0\}&$4q^{15}{+}2q^{16}$\\
\hline
4-1&\{110;01;0\}&$2q^{14}$\\
4-2&\{121;10;0\}&$8q^{14}$\\
4-3&\{112;21;0\}&$4q^{14}$\\
4-4&\{111;22;0\}&$4q^{14}$\\
4-5&\{211;11;2\}&$q^{12}$\\
4-6&\{110;02;0\}&$8q^{16}$\\
4-7&\{110;22;0\}&$8q^{15}{+}8q^{16}$\\
4-8&\{112;20;0\}&$4q^{16}$\\
4-9&\{120;02;0\}&$8q^{18}$\\
4-10&\{120;00;2\}&$16q^{17}{+}8q^{18}$\\
4-11&\{122;00;0\}&$4q^{17}$\\
4-12&\{210;01;0\}&$8q^{16}$\\
4-13&\{220;02;0\}&$4q^{18}{+}8q^{19}{+}2q^{20}$\\
4-14&\{222;00;0\}&$4q^{18}$\\
4-15&\{022;22;0\}&$q^{17}$\\
\end{tabular}
\end{ruledtabular}
\caption{\label{t:I}Bond lists and perimeter polynomials $D_c(q)$ for cluster
types of sizes $1{\le}n_c{\le}4$ in the \J1-\J2 model.
The labels for the cluster types are as in \Fig1.}
\end{table}

\begin{table*}
\renewcommand{\arraystretch}{1.25}
\begin{ruledtabular}
\begin{tabular}{ccc|ccc}
Cluster type, $c$&Bond List&$D_c(q)$&$c$&Bond List&$D_c(q)$\\\hline
5-1&\{0110;101;00;0\}&$2q^{16}$&5-23&\{1100;020;00;2\}&$16q^{19}{+}8q^{20}$\\
5-2&\{1120;010;01;0\}&$8q^{16}$&5-24&\{1100;220;02;0\}&$8q^{19}{+}4q^{20}$\\
5-3&\{1210;120;01;0\}&$8q^{16}$&5-25&\{1100;222;00;0\}&$8q^{18}$\\
5-4&\{2110;101;00;0\}&$4q^{16}$&5-26&\{1100;220;00;2\}&$8q^{18}{+}24q^{19}{+}8q^{20}$\\
5-5&\{1122;210;01;0\}&$4q^{16}$&5-27&\{1100;220;02;2\}&$4q^{17}$\\
5-6&\{1122;010;01;0\}&$4q^{15}{+}4q^{16}$&5-28&\{1120;202;00;0\}&$8q^{19}{+}8q^{20}$\\
5-7&\{1110;201;20;0\}&$8q^{16}$&5-29&\{1120;200;00;2\}&$8q^{19}{+}4q^{20}$\\
5-8&\{1110;221;00;0\}&$4q^{16}$&5-30&\{1022;200;00;0\}&$8q^{21}$\\
5-9&\{1112;201;20;0\}&$8q^{16}$&5-31&\{1200;002;20;0\}&$8q^{20}{+}24q^{21}{+}16q^{22}$\\
5-10&\{1121;212;10;0\}&$8q^{14}$&5-32&\{1200;000;20;2\}&$16q^{20}{+}32q^{21}{+}8q^{22}$\\
5-11&\{1111;022;22;0\}&$q^{16}$&5-33&\{1200;000;22;0\}&$24q^{20}$\\
5-12&\{1100;010;02;0\}&$8q^{18}$&5-34&\{1220;000;02;0\}&$8q^{20}{+}8q^{21}$\\
5-13&\{1102;210;00;0\}&$8q^{18}$&5-35&\{1022;000;22;0\}&$4q^{19}$\\
5-14&\{1120;010;02;0\}&$16q^{18}$&5-36&\{1010;200;01;0\}&$16q^{18}$\\
5-15&\{1210;100;02;0\}&$8q^{17}{+}8q^{18}$&5-37&\{1100;220;00;1\}&$16q^{17}{+}16q^{18}$\\
5-16&\{1120;210;02;0\}&$8q^{17}{+}8q^{18}$&5-38&\{2110;001;20;0\}&$8q^{18}$\\
5-17&\{1122;210;00;0\}&$8q^{17}$&5-39&\{0210;102;00;0\}&$32q^{20}$\\
5-18&\{1110;220;02;0\}&$8q^{17}{+}8q^{18}$&5-40&\{2102;010;00;0\}&$16q^{19}$\\
5-19&\{1110;222;00;0\}&$8q^{17}$&5-41&\{0102;012;00;0\}&$4q^{18}{+}12q^{19}{+}8q^{20}$\\
5-20&\{2112;110;20;0\}&$4q^{16}$&5-42&\{0220;202;00;0\}&$4q^{21}{+}16q^{22}{+}12q^{23}{+}2q^{24}$\\
5-21&\{1100;020;02;0\}&$8q^{20}$&5-43&\{2022;200;00;0\}&$8q^{21}{+}12q^{22}$\\
5-22&\{1100;022;00;0\}&$4q^{19}$&5-44&\{0222;220;00;0\}&$8q^{20}$\\
&&&5-45&\{2222;000;00;0\}&$q^{20}$\\
\end{tabular}
\end{ruledtabular}
\caption{\label{t:II}Bond lists and perimeter polynomials, $D_c(q)$, for the
quintet types $(n_c{=}5)$ of the \J1-\J2 model.
The labels for the cluster types are as in \Fig2.}
\end{table*}

Tables~\ref{t:III} and \ref{t:IV} in this Appendix give the bond lists, and the
perimeter Polynomials, $D_c(q)$, for the cluster types of the \J1-\J3 model.
The cluster types $c$  are those shown in \Figs{3}{4}.
\begin{table}
\renewcommand{\arraystretch}{1.25}
\begin{ruledtabular}
\begin{tabular}{ccc}
Cluster type, $c$&Bond List&$D_c(q)$\\\hline
1-1&\{\}&$q^{8}$\\
\hline
2-1&\{1\}&$2q^{12}$\\
2-2&\{3\}&$2q^{13}$\\
\hline
3-1&\{11;3\}&$2q^{16}$\\
3-2&\{11;0\}&$4q^{15}$\\
3-3&\{13;0\}&$8q^{16}{+}4q^{17}$\\
3-4&\{33;0\}&$4q^{17}{+}2q^{18}$\\
\hline
4-1&\{113;31;0\}&$2q^{20}$\\
4-2&\{131;10;0\}&$8q^{18}$\\
4-3&\{110;01;0\}&$4q^{18}$\\
4-4&\{111;30;0\}&$4q^{18}$\\
4-5&\{011;11;0\}&$q^{16}$\\
4-6&\{110;33;0\}&$8q^{19}{+}4q^{21}$\\
4-7&\{113;30;0\}&$4q^{19}$\\
4-8&\{110;03;0\}&$16q^{19}{+}8q^{20}$\\
4-9&\{113;00;0\}&$8q^{19}$\\
4-10&\{130;03;0\}&$4q^{20}{+}8q^{21}{+}2q^{22}$\\
4-11&\{130;00;3\}&$16q^{20}{+}16q^{21}{+}4q^{22}$\\
4-12&\{133;00;0\}&$12q^{20}$\\
4-13&\{130;03;1\}&$2q^{18}$\\
4-14&\{310;01;0\}&$4q^{19}{+}8q^{20}{+}2q^{21}$\\
4-15&\{330;03;0\}&$4q^{21}{+}8q^{22}{+}2q^{23}$\\
4-16&\{333;00;0\}&$4q^{21}$\\
4-17&\{033;33;0\}&$q^{20}$\\
\end{tabular}
\end{ruledtabular}
\caption{\label{t:III} Bond lists and perimeter polynomials, $D_c(q)$, for
cluster types of the \J1-\J3 model that have sizes $1{\le}n_c{\le}4$.
The labels for the cluster types are as in \Fig3.}
\end{table}
\begin{table*}
\renewcommand{\arraystretch}{1.25}
\begin{ruledtabular}
\begin{tabular}{ccc|ccc}
Cluster Type, $c$&Bond List&$D_c(q)$&$c$&Bond List&$D_c(q)$\\\hline
5-1&\{1133;310;01;0\}&$2q^{24}$&5-30&\{1130;303;00;0\}&$8q^{22}{+}8q^{24}$\\
5-2&\{1310;130;01;0\}&$8q^{22}$&5-31&\{1133;300;00;0\}&$2q^{22}$\\
5-3&\{1130;010;01;0\}&$8q^{21}$&5-32&\{1100;030;00;3\}&$8q^{22}{+}24q^{23}{+}32q^{24}{+}8q^{25}$\\
5-4&\{1133;010;01;0\}&$4q^{20}$&5-33&\{1100;030;03;0\}&$4q^{22}{+}20q^{23}{+}8q^{24}{+}4q^{25}$\\
5-5&\{0110;101;00;0\}&$4q^{20}$&5-34&\{1100;033;00;0\}&$24q^{23}$\\
5-6&\{3110;101;00;3\}&$4q^{20}$&5-35&\{1130;000;00;3\}&$16q^{23}{+}8q^{24}$\\
5-7&\{3110;101;00;0\}&$4q^{20}$&5-36&\{1130;003;00;0\}&$24q^{23}{+}16q^{24}$\\
5-8&\{1131;310;00;0\}&$8q^{21}$&5-37&\{1133;000;00;0\}&$4q^{22}$\\
5-9&\{1311;100;00;3\}&$4q^{20}$&5-38&\{1300;000;30;3\}&$16q^{24}{+}40q^{25}{+}24q^{26}{+}4q^{27}$\\
5-10&\{1110;301;00;0\}&$8q^{20}$&5-39&\{1300;000;33;0\}&$24q^{24}{+}12q^{25}$\\
5-11&\{1101;013;10;0\}&$8q^{19}$&5-40&\{1300;003;30;0\}&$32q^{24}{+}24q^{25}{+}24q^{26}{+}4q^{27}$\\
5-12&\{1111;300;00;3\}&$q^{20}$&5-41&\{1033;300;00;0\}&$8q^{24}{+}12q^{25}$\\
5-13&\{1130;310;03;0\}&$8q^{23}{+}4q^{25}$&5-42&\{1330;000;03;0\}&$32q^{24}{+}24q^{25}$\\
5-14&\{1133;310;00;0\}&$8q^{22}$&5-43&\{1333;000;00;0\}&$4q^{23}$\\
5-15&\{1310;100;03;0\}&$16q^{21}{+}8q^{23}$&5-44&\{1033;000;33;0\}&$8q^{23}$\\
5-16&\{1130;010;03;0\}&$8q^{22}{+}8q^{23}$&5-45&\{1100;330;00;1\}&$8q^{22}{+}8q^{23}{+}8q^{24}{+}4q^{25}$\\
5-17&\{1103;310;00;0\}&$16q^{21}$&5-46&\{3110;001;30;0\}&$4q^{23}$\\
5-18&\{1313;100;00;0\}&$8q^{21}{+}8q^{22}$&5-47&\{1130;303;00;1\}&$8q^{21}$\\
5-19&\{1100;010;03;0\}&$8q^{21}{+}8q^{22}{+}8q^{23}$&5-48&\{1010;300;01;0\}&$8q^{21}{+}24q^{22}{+}16q^{23}{+}8q^{24}$\\
5-20&\{1103;010;00;0\}&$16q^{22}$&5-49&\{3011;100;00;0\}&$8q^{22}{+}8q^{23}$\\
5-21&\{1110;303;00;0\}&$16q^{21}{+}8q^{23}$&5-50&\{1301;030;10;0\}&$8q^{21}$\\
5-22&\{1110;003;30;0\}&$8q^{22}{+}4q^{23}$&5-51&\{0103;013;00;0\}&$4q^{22}{+}20q^{23}{+}16q^{24}{+}12q^{25}{+}2q^{26}$\\
5-23&\{1113;300;00;0\}&$4q^{21}$&5-52&\{0310;103;00;0\}&$8q^{23}{+}32q^{24}{+}24q^{25}{+}4q^{26}$\\
5-24&\{0113;110;00;0\}&$8q^{20}$&5-53&\{1303;030;10;0\}&$16q^{22}$\\
5-25&\{1100;330;03;0\}&$4q^{22}{+}8q^{24}{+}2q^{26}$&5-54&\{3103;010;00;0\}&$32q^{23}{+}24q^{24}$\\
5-26&\{1100;330;00;3\}&$8q^{23}{+}8q^{24}{+}8q^{25}{+}4q^{26}$&5-55&\{0330;303;00;0\}&$4q^{25}{+}16q^{26}{+}12q^{27}{+}2q^{28}$\\
5-27&\{1100;333;00;0\}&$4q^{22}{+}8q^{23}$&5-56&\{3033;300;00;0\}&$8q^{25}{+}12q^{26}$\\
5-28&\{1100;330;03;3\}&$4q^{22}$&5-57&\{0333;330;00;0\}&$8q^{24}$\\
5-29&\{1130;300;00;3\}&$8q^{23}{+}4q^{24}$&5-58&\{3333;000;00;0\}&$q^{24}$\\
\end{tabular}
\end{ruledtabular}
\caption{\label{t:IV}Bond lists and perimeter polynomials, $D_c(q)$, for
quintet types $(n_c{=}5)$ of the \J1-\J3 model.
The labels for the cluster types are as in \Fig4.}
\end{table*}

\section{\label{a:relations}Relations between ``pure'' cluster types and cluster
types in the parent models}
Consider first the cluster types  of the \J1-\J2 model (\Figs{1}{2}).
At the bottom of these figures are the ``single'' (type 1-1) and the nine pure
\J2 cluster types.
Together, they are identical to the 10 cluster types of the (parent) \J2 model.
As pointed out in Appendix~\ref{a:iso} the cluster types of the \J2 model are
isomorphic to the cluster types of the \J1 model.
The pictorial representations of the cluster types of the \J1 model, shown
earlier in Fig.~3 of I, are therefore also applicable to the cluster types of the
\J2 model.

There are only four pure \J1 cluster types in the \J1-\J2 model.
Together with the single, they comprise only a subset of the 10 cluster types of
the parent \J1 model  (Fig.~3 of I).
The reason why some cluster types of the \J1 model are not among the pure \J1
cluster types of the \J1-\J2 model is the following.
As discussed in Sec.~\ref{sss:IIA3}, all cluster configurations that are present
in the \J1 model are also present in the \J1-\J2 model.
Some of these configurations have two \J1 bonds that are:
1) connected to the same spin, and
2) make a $90^\circ$ angle with each other.
In the \J1-\J2 model, each such configuration has at least one \J2 bond,
so that it is  a configuration of a mixed cluster of the \J1-\J2 model,
not a configuration of a pure \J1 cluster.
If  all configurations of a particular cluster type $c$  of the \J1 model
contain consecutive  \J1 bonds that make a $90^\circ$ angle, then this
cluster type is not among the pure \J1 cluster types of the  \J1-\J2 model.

Two \J1 bonds that are connected to the same spin are either at a $90^\circ$
angle or at a $180^\circ$ angle with each other.
Consider the pure \J1 cluster types that exist in the \J1-\J2 model.
Each such cluster type has only one configuration in this model.
All the spins in this configuration are on a single straight line.
In the parent \J1 model, on the other hand,  the same pure \J1 cluster type,
with the exception of cluster type 2-1, has several configurations.
For example, cluster type 3 in Fig.~3 of I has two configurations in the
\J1 model (configurations $3\alpha$ and $3\beta$ in Fig.~4 of I).
The same cluster type exists in the \J1-\J2 model as cluster type 3-1
(\Fig1 of the present paper), but the only configuration is $3\alpha$, with the
two \J1 bonds at $180^\circ$ angle.

Similar results are obtained for the pure cluster types in the \J1-\J3 model:
1) The pure \J3 cluster types of this model, together with the single, are
identical to the 10 cluster types of the parent \J3 model.
2) The pure \J1 cluster types, together with the single, are only a subset
of the cluster types of the parent \J1 model.
In this case, the reason why they are  only a subset is that in the \J1-\J3
model any configuration that has two consecutive \J1 bonds at $180^\circ$ angle,
also has at least one \J3 bond.
Therefore, in the \J1-\J3 model it is a configuration of a mixed cluster,
not of a pure \J1 cluster.
If all configurations of a particular cluster type $c$ of the \J1 model contain
consecutive \J1 bonds that make $180^\circ$ angle, then this cluster type is not
among the pure \J1 cluster types of  the \J1-\J3 model.

To follow up on the previous example of cluster type 3 in Fig.~3 of I, this
cluster type is identical to cluster type 3-2 in the  \J1-\J3 model
(\Fig3 of the present paper), which is a pure-\J1 cluster type.
Cluster type 3-2 of  the \J1-\J3 model has only one configuration, namely,
the $3\beta$ configuration in which the two consecutive \J1 bonds are at
$90^\circ$ angle.

\section{\label{a:comments}Comments on the probabilities of pure cluster types
in the \J1-\J2 and \J1-\J3 models}
As discussed in Appendix~\ref{a:relations}, the pure \J2 cluster types of the
\J1-\J2 model are identical to cluster types of the parent \J2 model.
The configurations for each of these cluster types are also the same in the two
models. However, the probability $P_c$ for each of these cluster types is lower
in the \J1-\J2 model than in the \J2 model.
The reason is that the \J1-\J2 perimeter for any configuration is larger than
the \J2 perimeter for the same configuration (see Sec.~\ref{ss:IIC}).
Similar remarks also apply to the probabilities for the pure \J3 cluster types
in the \J1-\J3 model, compared to the probabilities for the same cluster types
in the parent \J3 model.

The case of the pure \J1 cluster types is somewhat different.
Each of the pure \J1 cluster types in the \J1-\J2 model is identical to one of
the cluster types in the parent \J1 model.
However, the probability $P_c$ is lower in the \J1-\J2 model, for two reasons.
One is that, with the exception of cluster type 2-1, the number of
configurations in the \J1-\J2 model is smaller than in the parent \J1 model
(see Appendix~\ref{a:relations}).
The second reason is that for those configurations that exist in both models,
the \J1-\J2 perimeter is larger than the \J1 perimeter.
A larger perimeter implies a lower probability for the configuration.
For the same two reasons the probability $P_c$ for any pure \J1 cluster types in
the \J1-\J3 model is lower than that for the same cluster type in the \J1 model.


\begin{thebibliography}{18}
\bibitem[{not({\natexlab{a}})}]{note5}
\bibinfo{note}{A ``realization'' of a cluster type is any one cluster of this
  type. Similarly, a realization of a configuration is any one set of cation
  sites arranged in this configuration.}
\bibitem[{not({\natexlab{c}})}]{note10}
\bibinfo{note}{To avoid infinite values in the derivative, \dMdB was
  evaluated at a very low, but non-zero, temperature $T$.
}
\bibitem[{not({\natexlab{d}})}]{note11}
\bibinfo{note}{Including the derivative of $R(T,B)$ can lead to an error if
  $R(T,B)$ is obtained by the CQUIN's method. In this method the true clusters
  with $n_c{>}5$ are replaced by fictitious quintets. Including the derivative
  of $R(T,B)$ in the spectrum would lead to an artificial enhancement of
  spectral lines from quintets.}
\bibitem[{not({\natexlab{h}})}]{note:coincidence}
\bibinfo{note}{
In the range $J_2{/}J_1 {\le} 1$ the positions of the five spectral lines from
the 3-1 triplets remain exactly at $b_ 1 {=}7,9,11,13$ and $15$.
In the same range of \rJ2 the 3-2 triplets produce lines at exactly the
same fields, so thar all the 3-1 lines coincide with 3-2 lines.
However, when $\frac12{\le} J_2{/}J_1{\le}1$, the 3-2 triplets produce
additional lines at $b_1{<7}$.
The values of $b_1$ at these additonal lines depend on the ratio \rJ2.
}
\bibitem[{not({\natexlab{e}})}]{note7A}
\bibinfo{note}{Conclusions based on numerical simulations for specific values
of $S$ and specific ratios \rJ2 fall short of being proven.
However, the physical arguments in the following paper give strong additional support to
these conclusions.}
\bibitem[{not({\natexlab{f}})}]{note12}
\bibinfo{note}{All lattice sites of the square lattice in Fig.~1 of I can be
  reached from the central site by successive jumps between NN sites.
  The sublattice consisting of all sites that can be reached from the central
  site by successive jumps between 2nd neighbors is, by itself, also a square lattice.
  Compared to the original square lattice, this ``new'' square lattice has a
  lattice constant that is a factor of $\sqrt{2}$ larger, and it is rotated by
  $45^\circ$. The \J2 exchange bonds in the original square lattice are the NN
  exchange bonds in the new square lattice. All configurations, cluster types,
  cluster Hamiltonians, and cluster statistics for the original \J2 model are
  identical to those of the ``\J1 model'' for the new square lattice. A similar
  argument applies to the sublattice consisting of all sites that can be
  reached from the central site by successive jumps between 3rd neighbors.}
\bibitem[{not({\natexlab{g}})}]{note7}
\bibinfo{note}{In a given material the numerical value of $J$ is specified by
  the symmetry class of the neighbor associated with $J$, i.e., it depends on
  whether $J$ is \J1, or \J2, or \J3. However, if the material is not specified
  then the numerical value of $J$ is not known even if the symmetry class is
  known. In that case, $J$ is just a parameter in the exchange Hamiltonian. As
  long as it remains just a parameter, there is no difference between the
  exchange Hamiltonians of isomorphic cluster types.}
\end{thebibliography}
\newcommand{\noopsort}[1]{} \newcommand{\printfirst}[2]{#1}
  \newcommand{\singleletter}[1]{#1} \newcommand{\switchargs}[2]{#2#1}

\end{document}